\documentclass[prl,aps,twocolumn,epsfig,superscriptaddress,showpacs]{revtex4}
\usepackage{bbm}
\usepackage{amsfonts}
\usepackage[dvips]{graphicx}

\usepackage{mathrsfs}
\usepackage[intlimits]{amsmath}

\begin{document}

\author{Dong-Ling Deng}
 \affiliation{Theoretical Physics Division, Chern Institute of
Mathematics, Nankai University, Tianjin 300071, People's Republic of
China}
\author{Jing-Ling Chen}
 \email{chenjl@nankai.edu.cn}
\affiliation{Theoretical Physics Division, Chern Institute of
Mathematics, Nankai University, Tianjin 300071, People's Republic of
China}

\date{\today}

\title{Sufficient and Necessary Condition of
 Separability for Generalized Werner States}

\begin{abstract}
We introduce a sufficient and necessary condition for the
separability of a specific class of $N$ $d$-dimensional system
(qudits) states, namely  special generalized Werner state (SGWS):
$W^{[d^N]}(v)=(1-v)\frac{I^{(N)}}{d^N}+v|\psi
_d^N\rangle\langle\psi_d^N|$, where $
|\psi_d^N\rangle=\sum_{i=0}^{d-1}\alpha_i|i\cdots i\rangle$ is an
entangled pure state of $N$ qudits system and $\alpha_i$ satisfys
two restrictions: (i) $\sum_{i=0}^{d-1}\alpha_i\alpha_i^*=1$; (ii)
Matrix $\frac{1}{d}(I^{(1)}+\mathcal{T}\sum_{i\neq
j}\alpha_i|i\rangle\langle j|\alpha_j^*)$, where
$\mathcal{T}=\texttt{Min}_{i\neq j}\{1/|\alpha_i\alpha_j|\}$,  is a
density matrix. Our condition gives quite a simple and efficiently
computable way to judge whether a given SGWS is separable or not and
previously known separable conditions are shown to be special cases
of our approach.
\end{abstract}

 \maketitle

Entanglement and nonlocality are some of the most essential concepts
embodied in quantum mechanics~\cite{M.A. Nielsen}. A multiparticle
system is entangled if the states of this system cannot be prepared
locally by acting on the particles individually. Interest in
entangled systems has been heightened by proposed applications in
quantum computation and information, such as quantum parallelism
\cite{D.Deutsch}, quantum teleportation~\cite{C.
Bennett-G.Brassard,D.Bouwmeester}, dense coding~\cite{C.H.
Bennett-S.J.Wiesner,K. Mattle}, quantum cryptographic
schemes~\cite{A. Ekert}, entanglement swapping and remote state
preparation~\cite{M. A. Nielsen}, etc. There are many publications
that examined various aspects of entanglement~\cite{C.H.Bennett
D.P.Divincenzo, C.H.Bennett, M. A. Nielsen, W. Dur}, however,
entanglement is not yet fully understood
and many questions  remain open.\\
\indent One problem of great importance is to check whether a state,
generally mixed, is entangled or separable. Historically, violation
of Bell inequalities~\cite{J.S.Bell} is the first sufficient
condition of entanglement. If a  state violates any Bell inequality,
then it is certainly an entangled state. While the reverse is not
valid, there are various entangled  states that do not violate any
Bell inequality. Another brilliant step concerning this problem was
done by Peres, who proposed an necessary separability criterion
based on partial transpose of the composite density
operator~\cite{A.Peres}. Later on, Horodecki showed that this
criterion was a necessary and sufficient condition of separability
for the $2\otimes2$-or $2\otimes3$-dimensional states, but not a
sufficient condition any more for higher-dimensional states~\cite{M.
Horodecki-1,M. Horodecki-2}. There have been some other necessary
criteria for separability, such as reduction criterion~\cite{M.
Horodecki-P. Horodecki}, majorization criterion~\cite{M. A.
Nielsen-J. Kempe}, entanglement witnesses~\cite{B. Terhal},
realignment~\cite{O. Rudolph} and generalized realignment~\cite{S.
Albeverio}, as well as some necessary and sufficient operational
criteria for low rank density matrices~\cite{P. Horodecki-M.
Lewenstein-S. Albeverio}. More recently, some entanglement criteria
are also presented for continuous bipartite states~\cite{L.M.
Duan-E. Shchukin}. Nevertheless, up to now there is no general
necessary and sufficient condition for arbitrary states. Completely
solving the separability problem is far away from us.\\
\indent In this paper, we are interested in the separability
properties of $N$ particles $d$-dimensional ($N$ qudits) systems.
Their states are defined on finite dimensional Hilbert spaces
$H^{(N)}=H_1\otimes H_2\cdots \otimes H_N$, where $H_i$ denotes the
Hilbert space of the $i$th subsystem. A state in $H$ specified by a
density matrix $\rho$ is said to be fully separable if it is a
convex combination of tensor products:
\begin{eqnarray}
\rho=\sum_{\lambda}p_{\lambda}\rho_{\lambda}^{(1)}\otimes\cdots\otimes\rho_{\lambda}^{(N)},
\end{eqnarray}
where $0\leq p_{\lambda}\leq 1$, $\sum_{\lambda}p_{\lambda}=1$, and
$\rho_{\lambda}^{(k)}$ is a density matrix on $H_k$. In the
following, we will study the separability of a particular class of
$N$ qudits states, namely special generalized Werner state (SGWS).

The Werner state was originally defined in \cite{R.F.Werner} for two
qubits to show that some inseparable states admit the hidden
variable interpretation. The special generalized Werner state
(SGWS), $W^{[d^N]}(v)$  is defined as the convex combination of the
completely random state $\frac{1}{d^N}I^{(N)}$ and an entangled pure
state $|\psi_d^N\rangle\langle\psi_d^N|$,
\begin{eqnarray}\label{SGWS}
W^{[d^N]}(v)=(1-v)\frac{I^{(N)}}{d^N}+v|\psi_d^N\rangle\langle\psi_d^N|,
\end{eqnarray}
where $0\leq v\leq1$, $
|\psi_d^N\rangle=\sum_{i=0}^{d-1}\alpha_i|\tilde{i}\rangle$ (for
simple and convenient, we let $\tilde{i}$ denote  the repeated index
$i\cdots i$ and define
$|\tilde{i}\rangle=|i\rangle\otimes\cdots\otimes|i\rangle$) is an
entangled pure states of $N$ qudits system and $\alpha_i$ satisfys
two restrictions: (i) $\sum_{i=0}^{d-1}\alpha_i\alpha_i^*=1$; (ii)
Matrix $\varrho=\frac{1}{d}(I^{(1)}+\mathcal{T}\sum_{i\neq
j}\alpha_i|i\rangle\langle j|\alpha_j^*)$, where
$\mathcal{T}=\texttt{Min}_{i\neq j}\{1/|\alpha_i\alpha_j|\}$, is a
density matrix. The SGWS has been applied to study Bell's inequality
and local reality, and it has served as a test case of separability
arguments in a number of studies. As a result, studies of the
separability for SGWS itself is of great importance. In
~\cite{A.O.Pittenger-032313,A.O.Pittenger-042306,A.O.Pittenger-Opt},
Arthur O. Pittenger and Morton H. Rubin presented a necessary and
sufficient condition of separability for a special case, they proved
that, for
$|\psi_d^N\rangle=\frac{1}{\sqrt{d}}\sum_{i=0}^{d-1}|\tilde{i}\rangle$,
the SGWS is fully separable if and only if $v\leq(1+d^{n-1})^{-1}$.
Nevertheless, for general $|\psi_d^N\rangle$, no such result has
been known.
In this letter, we shall provide a necessary and sufficient
condition of separability for the SGWS. Our main result is as follows:\\
\indent \textit{Theorem} $1$: The special generalized Werner state
$W^{[d^N]}(v)$ defined in (\ref{SGWS}) is fully separable if and
only if $v\leq \frac{\mathcal {T}}{d^N+\mathcal{T}}$.

\textit{Proof}. We will take two steps to prove the above theorem.
In the first step, we shall prove that the condition is a  necessary
condition, in the second step, we prove the sufficiency part. The
method used here is similar to the method used in
ref.~\cite{A.O.Pittenger-Opt}.

Step $1$: In this part, we will prove that the condition is a
necessary condition. Typically, as mentioned above, the Peres
partial transpose condition is a general necessary condition for
separability, while it is not practical here to prove the necessity.
In \cite{A.O.Pittenger-042306,A.O.Pittenger-032313} a weaker but
useful condition for qudits was derived using the Cauchy-Schwarz
inequality. We will reproduce here their constriction for the sake
of completeness. Let $n=n_1\cdots n_N$ and $m=m_1\cdots m_N$ differ
in each component: $n_r\neq m_r$.  If a density matrix $\rho$ in
$H^{(N)}$ is separable, then the matrix elements of $\rho$ can be
written as
\begin{eqnarray}
\rho_{n,m}=\sum_{\lambda}
p_{\lambda}\prod_{r=1}^N\rho^{(r)}_{n_r,m_r}(\lambda).
\end{eqnarray}
Since each $\rho^{(r)}_{n_r,m_r}(\lambda)$ is a density matrix,
positivity requires that
$\sqrt{\rho^{(r)}_{n_r,n_r}(\lambda)}\sqrt{\rho^{(r)}_{m_r,m_r}(\lambda)}\geq|\rho^{(r)}_{n_r,m_r}(\lambda)|$
for each $r$ and $\lambda$. Then using the Cauchy-Schwarz inequality
we have:
\begin{eqnarray}\label{Nece-ineq}
&&\sqrt{\rho_{n_1\cdots n_N,n_1\cdots n_N}\rho_{m_1\cdots m_N,m_1\cdots m_N}}\nonumber\\
&&=\sqrt{\sum_{\lambda}p_{\lambda}\prod_{r=1}^N\rho^{(r)}_{n_r,n_r}(\lambda)}\sqrt{\sum_{\lambda}p_{\lambda}\prod_{r=1}^N\rho^{(r)}_{m_r,m_r}(\lambda)}\nonumber\\
&&\geq
\sum_{\lambda}p_{\lambda}\sqrt{\prod_{r=1}^N\rho^{(r)}_{n_r,n_r}(\lambda)\prod_{r=1}^N\rho^{(r)}_{m_r,m_r}(\lambda)}\nonumber\\
&&\geq\sum_{\lambda}p_{\lambda}\prod_{r=1}^N|\rho^{(r)}_{n_r,m_r}(\lambda)|
\geq|\rho_{\mu_1\cdots \mu_N,\nu_1\cdots \nu_N}|,
\end{eqnarray}
where, because of the Hermiticity of the density matrices,
$(\mu_r,\nu_r)$ may be either $(n_r,m_r)$ or $(m_r,n_r)$ for each
$r$. If the SGWS is fully separable, then use the
inequality~(\ref{Nece-ineq}) and choose $n$ and $m$ appropriately,
one can obtain the necessary condition:
\begin{eqnarray}
\frac{1-v}{d^N}\geq v|\alpha_i\alpha_j|.
\end{eqnarray}
For all $i$ and $j$ the above equality is valid, then we have
$\frac{1-v}{d^N}\geq v \times\texttt{Max}_{i\neq
j}\{|\alpha_i\alpha_j|\}$, i.e.,
$v\leq1/(d^N\times\texttt{Max}_{i\neq j}\{|\alpha_i\alpha_j|\}+1)$.
Not that $\mathcal{T}=\texttt{Min}_{i\neq
j}\{1/|\alpha_i\alpha_j|\}=1/\texttt{Max}_{i\neq
j}\{|\alpha_i\alpha_j|\}$, it is easy to get $v\leq \frac{\mathcal
{T}}{d^N+\mathcal{T}}$. Thus we have proved that the condition we
presented is a necessary condition of separability for the SGWS.

Step $2$: In this part, we will show that the condition is a
sufficient condition. We start with $v=\frac{\mathcal
{T}}{d^N+\mathcal{T}}$, then we prove that for all $v<
\frac{\mathcal {T}}{d^N+\mathcal{T}}$, the SGWS  is also separable.
For $v=\frac{\mathcal {T}}{d^N+\mathcal{T}}$, it is easy to rewrite
$W^{[d^N]}(v)$ as:
\begin{eqnarray}\label{w-1}
W^{[d^N]}(v)&&=\frac{\mathcal{T}}{d^N+\mathcal{T}}\sum_{i=0}^{d-1}\alpha_i|\tilde{i}\rangle\langle\tilde{i}|\alpha_i^*+\\
&&\frac{d^N}{d^N+\mathcal{T}}\left[\frac{1}{d^N}\left(I^{(N)}+\mathcal{T}\sum_{i\neq
j}^{d-1}\alpha_i|\tilde{i}\rangle\langle\tilde{j}|\alpha_j^*\right)\right].\nonumber
\end{eqnarray}
Since the first term in~(\ref{w-1}) is a sum of separable
projections, we only need to show that the  second term is also
separable. It is convenient in what follows to introduce a set of
fixed phase and to show that
\begin{eqnarray}\label{wT-2}
\rho^{(N)}=\frac{1}{d^N}\left(I^{(N)}+\mathcal{T}\sum_{i\neq
j}^{d-1}\alpha_i\xi_i|\tilde{i}\rangle\langle\tilde{j}|\alpha_j^*\xi_j^*\right),
\end{eqnarray}
where \{$|\xi_i|=1, i=0,\cdots,d-1$\}, is separable. We proceed by
induction. For $N=1$, we have
\begin{eqnarray}\label{rho1}
\rho^{(1)}=\frac{1}{d}\left(I^{(1)}+\mathcal{T}\sum_{i\neq
j}^{d-1}\alpha_i\xi_i|i\rangle\langle j|\alpha_j^*\xi_j^*\right).
\end{eqnarray}
Since $\mathcal{T}=\texttt{Min}_{i\neq j}\{1/|\alpha_i\alpha_j|\}$,
it follows that $|\rho_{i,j}|\leq 1/d$. From (\ref{rho1}) and the
definition of SGWS,
it is obviously that $\rho^{(1)}$ is a density matrix for any choice
of the parameters $\xi_i$.

Now assume that the density matrix of the form (\ref{wT-2}) is fully
separable for $N\geq2$, then we shall prove that it is fully
separable for $N+1$. Following the ideas in
\cite{A.O.Pittenger-Opt}, for a fixed choice of parameters $\xi_i$
define
\begin{eqnarray}
\mathbf{w}^{k}=(\xi_0z_0^{(k)},\cdots, \xi_{d-1}z^{(k)}_{d-1}),
\end{eqnarray}
where $z_i^{(k)}\in\{\pm1,\pm i\}$ and $\xi_i$ is independent of
$k$. We have a total of $4^d$ different vectors.
For each $\mathbf{w}^{(k)}$ define the product state
$\rho(k)=\rho^{(N)}(\mathbf{w}^{(k)})\otimes\rho^{(1)}(\mathbf{z}^{(k)})$
where $\mathbf{z}^{(k)}$ is equal to $\mathbf{w}^{(k)}$ with all the
$\xi_i$'s equal to $1$,
$\rho^{(N)}(\mathbf{w}^{(k)})=\frac{1}{d^N}\left(I^{(N)}+\mathcal{T}\sum_{i\neq
j}\alpha_i\xi_iz_i^{(k)}|\tilde{i}\rangle\langle\tilde{j}|\alpha_j^*\xi_j^*z_j^{(k)*}\right)$, 
and
$\rho^{(1)}(\mathbf{z}^{(k)})=\frac{1}{d}\left(I^{(1)}+\sum_{r\neq
s}z_r^{(k)}|r\rangle\langle s|z_s^{(k)*}\right)$. 
The states $\rho(k)$ are separable by the induction hypothesis,
consequently, the convex combination
$\rho^{(N+1)}=\frac{1}{4^d}\sum_k\rho(k)$ is also separable. Now we
multiply out the terms:
$\rho^{(N+1)}=\frac{1}{d^{N+1}}(I^{(N+1)}+I^{(N)}\otimes
F^{(1)}+F^{(N)}\otimes I^{(1)} +G)$
, here $F^{(1)}=\sum_{r\neq s}|r\rangle\langle s|\frac{1}{4^d}\sum_k
z_i^{(k)*}z_j^{(k)}$, 
$F^{(n)}=\sum_{i\neq j}\xi_i\alpha_i|\tilde{i}\rangle\langle
\tilde{j}|\alpha_j^*\xi_j^*\frac{1}{4^d}\sum_kz_i^{(k)}z_j^{(k)*}$, 
and $G=\sum_{i\neq j}\sum_{r\neq
s}\xi_i\xi_j^*\alpha_i\alpha_j^*|\tilde{i}\rangle\langle
\tilde{j}|\otimes |r\rangle\langle s|\frac{1}{4^d}
\times\sum_{k}z_i^{(k)}z_j^{(k)*}z_r^{(k)}z_s^{(k)*}$.
Since the components of $\mathbf{w}^{(k)}$ are chosen independently
of one another, $\sum_kz_r^{(k)}z_s^{(k)*}=0$ for $r\neq s$;
Consequently, $F^{(1)}$ and $F^{(n)}$ vanish and
$\frac{1}{4^d}\sum_{k}z_i^{(k)}z_j^{(k)*}z_r^{(k)}z_s^{(k)*}=\delta(i,r)\delta(j,s)$,
where $\delta(r,s)$ is the Kronecker delta function.
Then
\begin{eqnarray}
\rho^{(N+1)}=\frac{1}{d^{N+1}}\left[I^{(N+1)}+\mathcal{T}\sum_{i\neq
j}^{d-1}\alpha_i\alpha_j^*\xi_i|\tilde{i}i\rangle\langle\tilde{j}j|\xi_j^*\right].
\end{eqnarray}
which is the same form as (\ref{wT-2}) with $N\rightarrow N+1$, this
indicts that $\rho^{(N+1)}$ is also separable and consequently the
SGWS is separable when $v=\frac{\mathcal {T}}{d^N+\mathcal{T}}$ for all $N$. 
To complete the proof, there remains only one case to consider,
namely, $v< \frac{\mathcal {T}}{d^N+\mathcal{T}}$. Using the
provious results, it is very easy to prove the statement. For $v<
\frac{\mathcal {T}}{d^N+\mathcal{T}}$, we can rewrite the SGWS as :
\begin{widetext}
\begin{eqnarray}\label{vless}
W^{[d^N]}(v)&=&(1-v)\frac{I^{(N)}}{d^N}+v|\psi^N_d\rangle\langle\psi^N_d|\nonumber\\
&=&\left(1-\frac{v}{\frac{\mathcal{T}}{d^N+\mathcal{T}}}\right)\frac{I^{(N)}}{d^N}+
\frac{v}{\frac{\mathcal{T}}{d^N+\mathcal{T}}}\left[\left(1-\frac{\mathcal{T}}{d^N+\mathcal{T}}\right)
\frac{I^{(N)}}{d^N}+\frac{\mathcal{T}}{d^N+\mathcal{T}}|\psi^N_d\rangle\langle\psi^N_d|\right].
\end{eqnarray}
\end{widetext}
Obviously, the two parts of (\ref{vless}) are separable, thus the
convex combination of them is also separable. This complete the
proof of theorem.

Now returning to the  definition of SGWS (see equation
(\ref{SGWS})), we see that the restriction (ii) for $\alpha_i$ is
needed in the proof of the theorem. However, rough numerical results
show that, for some $\alpha_i$ not satisfying this restriction, the
theorem is also valid. Maybe we can rule out the restriction (ii) in
the definition of SGWS, while we do not know how to prove
analytically the theorem without it and we leave it as an open
question.

As an example, we present a class of SGWS: Let
$\alpha_0=\frac{1}{\sqrt{d}}\cos\theta$ and
$\alpha_i=\sqrt{\frac{d-\cos^2\theta}{d(d-1)}}$ ($i=1,\cdots, d-1$),
direct calculation shows that the  matrix
$\varrho=\frac{1}{d}(I^{(1)}+\mathcal{T}\sum_{i\neq
j}\alpha_i|i\rangle\langle j|\alpha_j^*)$ have $d-2$ zero
eigenvalues and two nonzero eigenvalues:
$\frac{1}{2}\left(1\pm\sqrt{\frac{(d-2)^2+(3d-4)\cos^2\theta}{d(d-\cos^2\theta)}}\right)$.
It follows that the matrix $\varrho$ is a density matrix and
consequently the corresponding  SGWS is separable if and only if
$v\leq \frac{d(d-1)}{d^N(d-\cos^2\theta)+d(d-1)}$. If we set
$\theta=0$, then $\alpha_i=\frac{1}{\sqrt{d}}$ and consequently the
corresponding  SGWS is separable if and only if $v\leq
\frac{1}{d^{N-1}+1}$, which is just the main result in
\cite{A.O.Pittenger-Opt}. Therefore, the necessary and sufficient
separable conditions presented in \cite{A.O.Pittenger-Opt} are shown
to be special cases of our theorem.


Our theorem also accord with other previous results of low
dimensional systems. Form the theorem,  for $d=2$,
$|\psi_2^N\rangle=\sin\theta|\tilde{0}\rangle+\cos\theta|\tilde{1}\rangle$,
we have $\mathcal{T}=1/\sin\theta \cos\theta$ and $W^{[2^N]}(v)$ is
fully separable if and only if $v\leq 1/(2^N\sin\theta
\cos\theta+1)$. On the other hand, for two qubits system, Wootters
presented a simple formula for the calculation of the entanglement
of formation~\cite{Wootters}
\begin{eqnarray}
E(\rho)=\mathcal
{E}(C(\rho))=h\left(\frac{1+\sqrt{1-C(\rho)^2}}{2}\right)
\end{eqnarray}
where $C(\rho)=$Max$\{0,\lambda_1-\lambda_2-\lambda_3-\lambda_4\}$
is the concurrence for two qubits density matrix,
$h(x)=-x$log$_2x-(1-x)$log$_2(1-x)$, and the $\lambda_i$s are the
eigenvalues, in decreasing order, of the Hermitian matrix
$R\equiv\sqrt{\sqrt{\rho}\tilde{\rho}\sqrt{\rho}}$. Since
$\mathcal{E}(C(\rho))$ is monotonically increasing and ranges from
$0$ to $1$ as $C(\rho)$ goes from $0$ to $1$, one can take the
concurrence as a measure of entanglement in its own right. For two
qubits Werner states under consideration $W^{[2^2]}$, direct
calculation show that
\begin{widetext}
\begin{eqnarray}\label{C-2qubits}
C(W^{[2^2]})&=&\texttt{Max}\{0,
\frac{1}{4}\sqrt{1+2v+v^2-4v^2\cos4\theta+4v\sin2\theta\sqrt{(1-v)^2+2v^2\cos4\theta}}\nonumber\\
&&-\sqrt{1+2v+v^2-4v^2\cos4\theta-4v\sin2\theta\sqrt{(1-v)^2+2v^2\cos4\theta}}-2+2v\}.
\end{eqnarray}
\end{widetext}
It is easy to check that $C(W^{[2^2]})=0$ if and only if $v\leq
1/(2^2\sin\theta \cos\theta+1)$, which is also obtained form the
theorem $1$. What's more, note that
$C(|\psi_2^2\rangle)=2\sin\theta\cos\theta$, one has that the
critical value of $v$ is monotonically  decreasing form $1$ to $1/3$
as $C(|\psi_2^2\rangle)=\sin\theta\cos\theta$ goes from $0$ to $1$.
This implies that  the critical value of $v$ is related to the
entanglement degree of the entangled pure state $|\psi_2^2\rangle$,
the greater the entanglement, the less the critical value of $v$.
While, surprisingly this is not valid for $d\geq3$, just as that the
maximal entangled states do not violate the Bell inequality
maximally for $d\geq3$~\cite{J. L. Chen}. From theorem $1$, if one
set two of $\alpha_i$s a little less than $1/\sqrt{2}$, then
$\mathcal{T}$ can be a value very close to $2$ and the critical
value of $v$ is close to $2/(d^N+2)$, which is smaller than
$d/(d^N+d)$ for the maximal entangled pure states
$|\psi_d^N\rangle=\frac{1}{\sqrt{d}}\sum_{i=0}^{d-1}|\tilde{i}\rangle$.
For instance, when $d=3$, $N=2$, choose
$|\psi_3^2\rangle=\frac{2}{3}|\tilde{0}\rangle+\frac{2}{3}|\tilde{1}\rangle+\frac{1}{3}|\tilde{2}\rangle$
which is not a maximal entangled states, one has that the
corresponding SGWS is separable if and only if $v\leq 1/5$, while
for
$|\psi_3^2\rangle=\frac{1}{\sqrt{3}}\sum_{i=0}^{d-1}|\tilde{i}\rangle$,
the corresponding SGWS is separable if and only if $v\leq 1/4$.

In summary, we have presented a sufficient and necessary condition
of separability for SGWS. Our condition gives quite a simple and
efficiently computable way to judge whether a given SGWS is
separable or not and previously known separable conditions are shown
to be special cases of our approach. Since the various use of SGWS
in quantum information, our results may be very useful for the study
of Bell inequalities, quantum entanglement measurement, distillation
protocols, etc.

This work was supported in part by NSF of China (Grant No.
10605013), Program for New Century Excellent Talents in University,
and the Project-sponsored by SRF for ROCS, SEM.

\end{document}